\documentclass{elsartNoFoot}

\usepackage{latin1}
\usepackage{amssymb}
\usepackage{url}

\usepackage[pdftex]{color}
\usepackage{jb}
\usepackage{graphicx}

\newcommand{\LABEL}[1]{\label{#1}}

\newcommand{\ra}{\rightarrow}
\renewcommand{\leq}{\leqslant}
\renewcommand{\geq}{\geqslant}

\newcommand{\tpl}[1]{\langle #1 \rangle}	
\newcommand{\set}[1]{\{ #1 \}}			

\newcommand{\G}{\mathcal{G}}
\newcommand{\R}{\mathcal{R}}
\newcommand{\ntEquiv}{\equiv}
\newcommand{\NT}{\mathcal{N}}

\newcommand{\LL}{\mathcal{L}}		
\newcommand{\OO}{\mathcal{O}}		

\renewcommand{\:}[4]{%
	{%
	\renewcommand{\:}[4]{%
		{%
		\renewcommand{\:}[4]{error\error}%
		\renewcommand{\j}{{##2}}%
		{##4}%
		##1...##1%
		\renewcommand{\j}{{##3}}%
		{##4}%
		}%
	}%
	\renewcommand{\i}{{#2}}%
	{#4}%
	#1\ldots#1%
	\renewcommand{\i}{{#3}}%
	{#4}%
	}%
}

\newcommand{\THM}[3]{%
	\vspace{2ex plus 3ex}%
	\begin{#1}%
	\Ite{#2}{ (#2) }{ }%
	{#3}%
	\end{#1}%
	\vspace{2ex plus 3ex}%
	}
\newcommand{\PRF}[2]{%
	\begin{pf}%
	{#2}%
	\end{pf}%
	\vspace{2ex plus 3ex}%
	}
\newcommand{\EXAMPLE}[2]   {\THM{example}   {#1}{#2}}
\newcommand{\LEMMA}[2]     {\THM{lemma}     {#1}{#2}}
\newcommand{\COROLLARY}[2] {\THM{corollary} {#1}{#2}}

\newcommand{\DEFINITION}[2]{\THM{definition}{#1}{#2}}

\newcommand{\PROOF}[1]     {\PRF{\em Proof. }       {#1}}

\begin{document}

\newsavebox{\atpic}
\savebox{\atpic}{\includegraphics[scale=0.08]{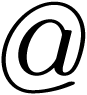}}

\setReceivedPrefix{}

\begin{frontmatter}

\title{Repetition-Free Derivability from a Regular Grammar is NP-Hard}
\author{Jochen Burghardt}
\address{jochen.burghardt\usebox{\atpic}alumni.tu-berlin.de}
\received{February 2016}

\begin{abstract}
We prove the NP-hardness of the problem
whether a given word can be derived from a given
regular grammar without repeated occurrence of any nonterminal.
\end{abstract}

\begin{keyword}
Regular grammar;
repetition-free derivation;
NP-hard
\end{keyword}

\end{frontmatter}

\section{Introduction}
\LABEL{Introduction}

Let a regular word grammar $\G$ be given.
We ask whether a given word $\omega$ can be derived from
$\G$ without repeated occurrence of any nonterminal.
We prove in Sect.~\ref{Repetition-free derivability}
that the problem of deciding this property is NP-hard
in general.
As a consequence, it is NP-hard also for all superclasses of regular
grammars, such as context-free, context-sensitive, and
unrestricted grammars.

In Sect.~\ref{Longest repetition-free derivable words},
we present some ideas to prove the NP-hardness of a related problem,
viz.\ of determining the length of the longest word repetition-free
derivable from a given grammar.
However, we didn't yet succeed in finding a proof for that claim.

In Sect.~\ref{Application to sequence guessing},
we present the original motivation of considering repetition-free
derivations, which was a rather particular problem from artificial
intelligence.

The problem of deciding repetition-free derivability
looks quite similar to that of deciding the existence of a
Hamiltonian path in a given undirected graph, which is well-known to
be NP-complete \cite[Thm.7.35, Sect.7.5, p.262]{Sipser.1997}.
However, both problems differ in
\begin{itemize}
\item presence of terminals/edge labels,
\item the set of nonterminals/nodes in a derivation/path 
	(arbitrary vs.\ full set), and
\item the admitted start and end nonterminals/nodes of a derivation/path
	(fixed start and end symbols vs.\ arbitrary nodes),
	respectively.
\end{itemize}
For this reason, a reduction of the Hamiltonian path problem to the
repetition-free derivability problem is not immediate obvious.

\section{Definitions}
\LABEL{Definitions}

\DEFINITION{Regular grammar}{%
\LABEL{Regular grammar}%
Following \cite[Sect.9.1/4.2, p.217/79]{Hopcroft.Ullman.1979},
a regular (word) grammar $\G$ is defined as a tuple
$\tpl{\NT,\Sigma,\R,S}$,
where $\NT$ and $\Sigma$ are disjoint finite sets of nonterminal and
terminal symbols, respectively, $S \in \NT$ is called the start
symbol, and $\R$ is a finite set of rules of the form $A ::= b C$
or $A ::= b$,
where $A, C \in \NT$ and $b \in \Sigma$.

A derivation from $\G$ is a finite sequence
$$\begin{array}{lll}
S & \ra & a_1 X_1	\\
	& \ra & a_1 a_2 \; X_2	\\
	& \ra & \ldots	\\
	& \ra & a_1 a_2 \ldots a_{n-1} \; X_{n-1}	\\
	& \ra & a_1 a_2 \ldots a_{n-1} a_n \; X_n	\\
	& \ra & a_1 a_2 \ldots a_{n-1} a_n a_{n+1}	\\
\end{array}$$
where $\:,1{n+1}{a_\i} \in \Sigma$ are terminal symbols,
$\:,1n{X_\i} \in \NT$ are nonterminal symbols,
and
$$\begin{array}{llll}
S & ::= a_1 \; X_1& ,	\\
X_1 & ::= a_2 \; X_2 & ,	\\
\ldots && ,	\\
X_{n-1} & ::= a_n \; X_n & , \mbox{ and}	\\
X_n & ::= a_{n_1}	\\
\end{array}$$
are rules from $\R$.
We say that the nonterminals $\:,1n{X_\i}$ occur in that derivation.
A word $\omega \in \Sigma^*$ is derivable from $\G$ is a derivation
$S \ra \ldots \ra \omega$ exists.
The language produced by $\G$ is denoted by $\LL(\G)$,
it is defined as the set of all words derivable from $\G$.
\qed
}

\DEFINITION{Conjunctive normal form formula}{%
\LABEL{Conjunctive normal form formula}%
Let a set $\set{ \:,1m{x_\i} }$ of propositional variables be given.
A boolean formula in (3-literal) conjunctive normal form
is given as a conjunction $\kappa = \:{\cdot}1n{\kappa_\i}$,
where the $j$.th conjunct $\kappa_j$ has the form
$y_{j1}+y_{j2}+y_{j3}$
and each literal $y_{jk}$ satistifes
$y_{jk} \in \set{ \:,1m{x_\i} } \cup \set{ \:,1m{\overline{x}_\i} }$.

Given an assignment of truth values $0$ or $1$ to the variables
$\:,1m{x_\i}$,
\begin{itemize}
\item a literal $x_i$ and $\overline{x}_i$ is satisfied if
	$1$ and $0$ has been assigned to $x_i$, respectively;
\item a conjunct $\kappa_j = y_{j1}+y_{j2}+y_{j3}$ is satisfied 
	if at least
	one of its literals $y_{j1}$, $y_{j2}$, $y_{j3}$ is; and
\item the whole formula 
	$\kappa = \:{\cdot}1n{\kappa_\i}$ is
	satisfied if each of its conjuncts $\kappa_j$ is.
\end{itemize}

The formula is called satisfiable if it is satisfied by some
assignment.
It is well-known that the problem of deciding the satistfiability of a
given 3-literal conjunctive normal form formula is NP-complete
(e.g.\ \cite[Sect.10.4, Thm.10.4, p.384]{Aho.Hopcroft.Ullman.1974}).
\qed
}


\section{Repetition-free derivability}
\LABEL{Repetition-free derivability}

The ordinary derivability problem for regular word grammars can be
solved within an time upper bound of $\OO(n \cdot s^2)$,
where $n$ and $s$ is the length of the input string and the number of
nonterminals, respectively 
\cite[Sect.4.3.3, p.153]{Hopcroft.Motwani.Ullman.2003}.\footnote{%
	Hopcroft et.\ al.\ explain their algorithm on nondeterministic
	finite automata, using the number of states for $s$.
	However, carrying-over to regular grammars is
	straight-forward.
}
In contrast, repetition-free derivability is NP-hard, as we show in
the following.

We reduce the satisfiability problem for conjunctive normal forms,
which is well-known to be NP-complete 
\cite[Thm.10.3, Sect.10.4, p.379]{Aho.Hopcroft.Ullman.1974},
to the repetition-free derivability problem.
We give the mapping of a former to a latter problem in 
Def.~\ref{Grammar corresponding to a conjunctive normal form},
and prove it a reduction in
Cor.~\ref{Repetition-Free Derivability from a Regular Grammar is NP-Hard},
based essentially on Lem.~\ref{Repetition-Free derivability}.

\begin{figure}
$$\begin{array}{lcl@{\hspace*{0.5cm}}l}
\hline
S_{i-1} & ::= & a \; X_{i1}
	& \mbox{for } i=\:,1m\i	\\
S_{i-1} & ::= & a \; \overline{X}_{i1}
	& \mbox{for } i=\:,1m\i	\\
X_{ij} & ::= & a \; X_{i,j+1}
	& \mbox{for } i=\:,1m\i \mbox{ and } j=\:,1{n-1}\i	\\
\overline{X}_{ij} & ::= & a \; \overline{X}_{i,j+1}
	& \mbox{for } i=\:,1m\i \mbox{ and } j=\:,1{n-1}\i	\\
X_{in} & ::= & a \; S_i & \mbox{for } i=\:,1m\i	\\
\overline{X}_{in} & ::= & a \; S_i 
	& \mbox{for } i=\:,1m\i	\\
S_m & ::= & b T_0	\\
\hline
T_{j-1} & ::= & c \; \gamma_{jk} 
	& \mbox{for } j=\:,1n\i	
	\mbox{ and } k=1,2,3	\\
\gamma_{jk} & ::= & e \; T_j
	& \mbox{for } j=\:,1n\i
	\mbox{ and } k=1,2,3	\\
T_n & ::= & d	\\
\hline
\multicolumn{4}{l}{\mbox{where the mapping } \gamma 
	\mbox{ is defined by}}	\\
\gamma_{jk} & = & X_{ij} & \mbox{for } y_{jk} = x_i	\\
\gamma_{jk} & = & \overline{X}_{ij} 
	& \mbox{for } y_{jk} = \overline{x}_i	\\
\hline
\end{array}$$
\caption{Grammar rules in 
	Def.~\ref{Grammar corresponding to a conjunctive normal form}}
\label{Grammar rules in Def.}
\end{figure}

\DEFINITION{Grammar corresponding to a conjunctive normal form}{%
\label{Grammar corresponding to a conjunctive normal form}%
Given a conjunctive normal form formula
as in Def.~\ref{Conjunctive normal form formula},
we define a ``corresponding''
a regular grammar $\G = \tpl{\NT,\Sigma,\R,S_0}$
as follows.

Let
$\NT
=    \set{ \:,0m{S_\i}, \:,0n{T_\i} }
\cup \set{ X_{ij}, \overline{X}_{ij}
	\mid 1 \leq i \leq m \land 1 \leq j \leq n }$
be the set of nonterminal symbols,
let
$\Sigma = \set{a,b,c,d}$
be the set of terminal symbols.
Let the rules $\R$ be as shown in Fig~\ref{Grammar rules in Def.}.
We refer to the topmost 7 and the next 3 lines as the upper and lower
grammar part, respectively.
\qed
}

{
\definecolor{coEqvA}	{rgb}{0.80,0.00,0.00}
\definecolor{coEqvB}	{rgb}{0.40,0.00,0.00}
\definecolor{coEqvC}	{rgb}{0.00,0.80,0.00}
\definecolor{coEqvD}	{rgb}{0.00,0.40,0.00}
\definecolor{coEqvE}	{rgb}{0.80,0.00,0.80}
\definecolor{coEqvF}	{rgb}{0.40,0.00,0.40}
\definecolor{coEqvG}	{rgb}{0.00,0.80,0.80}
\definecolor{coEqvH}	{rgb}{0.00,0.40,0.40}
\newcommand{\0}[1]{\textcolor{coEqvA}{#1}}
\newcommand{\1}[1]{\textcolor{coEqvB}{#1}}
\newcommand{\2}[1]{\textcolor{coEqvC}{#1}}
\newcommand{\3}[1]{\textcolor{coEqvD}{#1}}
\newcommand{\4}[1]{\textcolor{coEqvE}{#1}}
\newcommand{\5}[1]{\textcolor{coEqvF}{#1}}
\newcommand{\6}[1]{\textcolor{coEqvG}{#1}}
\newcommand{\7}[1]{\textcolor{coEqvH}{#1}}

\begin{figure}
$$\begin{array}{l@{\hspace*{1cm}}l@{\hspace*{1cm}}l}
\begin{array}[t]{l@{\;}c@{\;}l@{\;}c@{\;}l@{\;}c@{\;}l}
S_0 & ::= & a \0{X_{11}}
	& \mid & a \1{\overline{X}_{11}}	\\
\0{X_{11}} & ::= & a \0{X_{12}}
	& \mid & e T_1	\\
\0{X_{12}} & ::= & a \0{X_{13}}	\\
\0{X_{13}} & ::= & a S_1	\\
S_1 & ::= & a \2{X_{21}}
	& \mid & a \3{\overline{X}_{21}}	\\
\2{X_{21}} & ::= & a \2{X_{22}}	\\
\2{X_{22}} & ::= & a \2{X_{23}}
	& \mid & e T_2	\\
\2{X_{23}} & ::= & a S_2	\\
S_2 & ::= & a \4{X_{31}}
	& \mid & a \5{\overline{X}_{31}}	\\
\4{X_{31}} & ::= & a \4{X_{32}} &&	\\
\4{X_{32}} & ::= & a \4{X_{33}}
	& \mid & e T_2	\\
\4{X_{33}} & ::= & a S_3 &&	\\
S_3 & ::= & a \6{X_{41}}
	& \mid & a \7{\overline{X}_{41}}	\\
\6{X_{41}} & ::= & a \6{X_{42}}
	& \mid & e T_1	\\
\6{X_{42}} & ::= & a \6{X_{43}} &&	\\
\6{X_{43}} & ::= & a S_4
	& \mid & e T_3	\\
S_4 & ::= & b T_0	\\
\end{array}
&
\begin{array}[t]{l@{\;}c@{\;}l@{\;}c@{\;}l}
\\
\1{\overline{X}_{11}} & ::= & a \1{\overline{X}_{12}}	\\
\1{\overline{X}_{12}} & ::= & a \1{\overline{X}_{13}}	\\
\1{\overline{X}_{13}} & ::= & a S_1
	& \mid & e T_3	\\
\\
\3{\overline{X}_{21}} & ::= & a \3{\overline{X}_{22}}	
	& \mid & e T_1	\\
\3{\overline{X}_{22}} & ::= & a \3{\overline{X}_{23}}	\\
\3{\overline{X}_{23}} & ::= & a S_2
	& \mid & e T_3	\\
\\
\5{\overline{X}_{31}} & ::= & a \5{\overline{X}_{32}} &&	\\
\5{\overline{X}_{32}} & ::= & a \5{\overline{X}_{33}} &&	\\
\5{\overline{X}_{33}} & ::= & a S_3 &&	\\
\\
\7{\overline{X}_{41}} & ::= & a \7{\overline{X}_{42}} &&	\\
\7{\overline{X}_{42}} & ::= & a \7{\overline{X}_{43}}
	& \mid & e T_2	\\
\7{\overline{X}_{43}} & ::= & a S_4 &&	\\
\end{array}
&
\begin{array}[t]{|l@{\;}c@{\;}l@{\;}c@{\;}l@{\;}c@{\;}l}
\multicolumn{1}{l}{}	\\
\multicolumn{1}{l}{}	\\
\multicolumn{1}{l}{}	\\
\multicolumn{1}{l}{}	\\
\multicolumn{1}{l}{}	\\
\multicolumn{1}{l}{}	\\
\multicolumn{1}{l}{}	\\
\multicolumn{1}{l}{}	\\
\multicolumn{1}{l}{}	\\
\multicolumn{1}{l}{}	\\
\multicolumn{1}{l}{}	\\
\multicolumn{1}{l}{}	\\
\multicolumn{1}{l}{}	\\
\hline
T_0 & ::= & c \0{X_{11}} 
	& \mid & c \3{\overline{X}_{21}}
	& \mid & c \6{X_{41}}	\\
T_1 & ::= & c \2{X_{22}} 
	& \mid & c \4{X_{32}}
	& \mid & c \7{\overline{X}_{42}}	\\
T_2 & ::= & c \1{\overline{X}_{13}} 
	& \mid & c \3{\overline{X}_{23}}
	& \mid & c \6{X_{43}}	\\
T_3 & ::= & d &&&&	\\
\end{array}
\\
\end{array}$$
\caption{Example grammar in 
	Exm.~\ref{Conjunctive normal form and corresponding grammar}}
\label{Example grammar in Exm.}
\end{figure}

\begin{figure}
\begin{picture}(150,40)
\put(3.000,30.000){\makebox(0.000,0.000){$S_0$}}
\put(5.000,30.000){\vector(1,1){4.000}}
\put(5.000,30.000){\vector(1,-1){4.000}}

\put(12.000,35.000){\makebox(0.000,0.000){$\0{X_{11}}$}}
\put(12.000,25.000){\makebox(0.000,0.000){$\1{\overline{X}_{11}}$}}
\put(15.000,35.000){\vector(1,0){2.000}}
\put(15.000,25.000){\vector(1,0){2.000}}
\put(20.000,35.000){\makebox(0.000,0.000){$\0{X_{12}}$}}
\put(20.000,25.000){\makebox(0.000,0.000){$\1{\overline{X}_{12}}$}}
\put(23.000,35.000){\vector(1,0){2.000}}
\put(23.000,25.000){\vector(1,0){2.000}}
\put(28.000,35.000){\makebox(0.000,0.000){$\0{X_{13}}$}}
\put(28.000,25.000){\makebox(0.000,0.000){$\1{\overline{X}_{13}}$}}
\put(31.000,34.000){\vector(1,-1){4.000}}
\put(31.000,26.000){\vector(1,1){4.000}}

\put(37.000,30.000){\makebox(0.000,0.000){$S_1$}}
\put(39.000,30.000){\vector(1,1){4.000}}
\put(39.000,30.000){\vector(1,-1){4.000}}
\put(46.000,35.000){\makebox(0.000,0.000){$\2{X_{21}}$}}
\put(46.000,25.000){\makebox(0.000,0.000){$\3{\overline{X}_{21}}$}}
\put(49.000,35.000){\vector(1,0){2.000}}
\put(49.000,25.000){\vector(1,0){2.000}}
\put(54.000,35.000){\makebox(0.000,0.000){$\2{X_{22}}$}}
\put(54.000,25.000){\makebox(0.000,0.000){$\3{\overline{X}_{22}}$}}
\put(57.000,35.000){\vector(1,0){2.000}}
\put(57.000,25.000){\vector(1,0){2.000}}
\put(62.000,35.000){\makebox(0.000,0.000){$\2{X_{23}}$}}
\put(62.000,25.000){\makebox(0.000,0.000){$\3{\overline{X}_{23}}$}}
\put(65.000,34.000){\vector(1,-1){4.000}}
\put(65.000,26.000){\vector(1,1){4.000}}

\put(71.000,30.000){\makebox(0.000,0.000){$S_2$}}
\put(73.000,30.000){\vector(1,1){4.000}}
\put(73.000,30.000){\vector(1,-1){4.000}}
\put(80.000,35.000){\makebox(0.000,0.000){$\4{X_{31}}$}}
\put(80.000,25.000){\makebox(0.000,0.000){$\5{\overline{X}_{31}}$}}
\put(83.000,35.000){\vector(1,0){2.000}}
\put(83.000,25.000){\vector(1,0){2.000}}
\put(88.000,35.000){\makebox(0.000,0.000){$\4{X_{32}}$}}
\put(88.000,25.000){\makebox(0.000,0.000){$\5{\overline{X}_{32}}$}}
\put(91.000,35.000){\vector(1,0){2.000}}
\put(91.000,25.000){\vector(1,0){2.000}}
\put(96.000,35.000){\makebox(0.000,0.000){$\4{X_{33}}$}}
\put(96.000,25.000){\makebox(0.000,0.000){$\5{\overline{X}_{33}}$}}
\put(99.000,34.000){\vector(1,-1){4.000}}
\put(99.000,26.000){\vector(1,1){4.000}}

\put(105.000,30.000){\makebox(0.000,0.000){$S_3$}}
\put(107.000,30.000){\vector(1,1){4.000}}
\put(107.000,30.000){\vector(1,-1){4.000}}
\put(114.000,35.000){\makebox(0.000,0.000){$\6{X_{41}}$}}
\put(114.000,25.000){\makebox(0.000,0.000){$\7{\overline{X}_{41}}$}}
\put(117.000,35.000){\vector(1,0){2.000}}
\put(117.000,25.000){\vector(1,0){2.000}}
\put(122.000,35.000){\makebox(0.000,0.000){$\6{X_{42}}$}}
\put(122.000,25.000){\makebox(0.000,0.000){$\7{\overline{X}_{42}}$}}
\put(125.000,35.000){\vector(1,0){2.000}}
\put(125.000,25.000){\vector(1,0){2.000}}
\put(130.000,35.000){\makebox(0.000,0.000){$\6{X_{43}}$}}
\put(130.000,25.000){\makebox(0.000,0.000){$\7{\overline{X}_{43}}$}}
\put(133.000,34.000){\vector(1,-1){4.000}}
\put(133.000,26.000){\vector(1,1){4.000}}

\put(139.000,30.000){\makebox(0.000,0.000){$S_4$}}

\put(28.000,10.000){\makebox(0.000,0.000)[r]{$\ldots$}}
\put(33.000,10.000){\makebox(0.000,0.000){$S_4$}}
\put(36.000,10.000){\vector(1,0){5.000}}
\put(43.000,10.000){\makebox(0.000,0.000){$T_0$}}
\put(46.000,10.000){\vector(2,1){10.000}}
\put(46.000,10.000){\vector(1,0){10.000}}
\put(46.000,10.000){\vector(2,-1){10.000}}
\put(59.000,15.000){\makebox(0.000,0.000){$\0{X_{11}}$}}
\put(59.000,10.000){\makebox(0.000,0.000){$\3{\overline{X}_{21}}$}}
\put(59.000,5.000){\makebox(0.000,0.000){$\6{X_{41}}$}}
\put(62.000,15.000){\vector(2,-1){10.000}}
\put(62.000,10.000){\vector(1,0){10.000}}
\put(62.000,5.000){\vector(2,1){10.000}}
\put(75.000,10.000){\makebox(0.000,0.000){$T_1$}}
\put(78.000,10.000){\vector(2,1){10.000}}
\put(78.000,10.000){\vector(1,0){10.000}}
\put(78.000,10.000){\vector(2,-1){10.000}}
\put(91.000,15.000){\makebox(0.000,0.000){$\2{X_{22}}$}}
\put(91.000,10.000){\makebox(0.000,0.000){$\4{X_{32}}$}}
\put(91.000,5.000){\makebox(0.000,0.000){$\7{\overline{X}_{42}}$}}
\put(94.000,15.000){\vector(2,-1){10.000}}
\put(94.000,10.000){\vector(1,0){10.000}}
\put(94.000,5.000){\vector(2,1){10.000}}
\put(107.000,10.000){\makebox(0.000,0.000){$T_2$}}
\put(110.000,10.000){\vector(2,1){10.000}}
\put(110.000,10.000){\vector(1,0){10.000}}
\put(110.000,10.000){\vector(2,-1){10.000}}
\put(123.000,15.000){\makebox(0.000,0.000){$\1{\overline{X}_{13}}$}}
\put(123.000,10.000){\makebox(0.000,0.000){$\3{\overline{X}_{23}}$}}
\put(123.000,5.000){\makebox(0.000,0.000){$\6{X_{43}}$}}
\put(126.000,15.000){\vector(2,-1){10.000}}
\put(126.000,10.000){\vector(1,0){10.000}}
\put(126.000,5.000){\vector(2,1){10.000}}
\put(139.000,10.000){\makebox(0.000,0.000){$T_3$}}

\end{picture}
\caption{Illustration of example grammar in 
	Exm.~\ref{Conjunctive normal form and corresponding grammar}}
\label{Illustration of example grammar in Exm.}
\end{figure}

\EXAMPLE{Conjunctive normal form and corresponding grammar}{
\LABEL{Conjunctive normal form and corresponding grammar}%
As an example, the conjunctive normal form
$$(\0{x_1}+\3{\overline{x}_2}+\6{x_4})
\cdot (\2{x_2}+\4{x_3}+\7{\overline{x}_4})
\cdot (\1{\overline{x}_1}+\3{\overline{x}_2}+\6{x_4})$$
corresponds to the grammar shown in Fig.~\ref{Example grammar in
Exm.},
where different colors indicate different variables, while light and
dark shades indicate unnegated and negated occurrences, respectively.
The $S_j ::= \ldots$ rules of the lower part are shown bottom
right,
its $\gamma_{j+1,k} ::= \ldots$ are integrated as alternatives in
the upper part's rules.
See also the illustration in 
Fig.~\ref{Illustration of example grammar in Exm.}, where upper and
lower part are strictly separated, and their common nonterminals 
(like $\0{X_{11}}$) are shown twice.
Observe that no nonterminal occurs multiply in the upper part alone, and
likewise none does in the lower.
\qed
}


\begin{figure}
\begin{center}
$$\begin{array}{|cc||c|ccc|}
\hline
y_{jk} 
	& x_i 
	& \mbox{sat} 
	& \mbox{initial} 
	& \mbox{final} 
	& \mbox{rep}	\\
\hline
x_i 
	& 0 
	& -
	& X_{ij} 
	& X_{ij} 
	& +	\\
\overline{x}_i 
	& 0 
	& +
	& X_{ij} 
	& \overline{X}_{ij} 
	& -	\\
x_i 
	& 1 
	& +
	& \overline{X}_{ij} 
	& X_{ij} 
	& -	\\
\overline{x}_i 
	& 1 
	& -
	& \overline{X}_{ij} 
	& \overline{X}_{ij} 
	& +	\\
\hline
\end{array}$$
\caption{Satisfied literal vs.\ repetition-free $ce$ derivation in
Lem.~\ref{Repetition-Free Derivability from a Regular Grammar is NP-Hard}}
\LABEL{Satisfied literal vs. repetition-free $ce$ derivation in Lem.}
\end{center}
\end{figure}

\LEMMA{Repetition-Free derivability}{%
\label{Repetition-Free derivability}%
Given a conjunctive normal form formula $\kappa$
as in Def.~\ref{Conjunctive normal form formula},
and its corresponding grammar $\G$ as in 
Def.~\ref{Grammar corresponding to a conjunctive normal form},
the word $\omega = a^{(n+1) \cdot m} b (ce)^n d$
has a repetition-free derivation from $\G$
iff $\kappa$ has a satisfying variable assignment.
}
\PROOF{
First, note that
symbols $a$ and $b$ are only produced by the upper grammar part;
similarly, symbols $c$ and $d$ are only produced by the lower one.
Therefore,
in order to derive a word starting with $a^{(n+1) \cdot m} b$,
the rules of the upper grammar part must be applied $(n+1) \cdot m + 1$ 
times, leading to an initial derivation part 
$S_0 
	\stackrel{*}{\ra} a^{(n+1) \cdot m} S_m 
	\ra a^{(n+1) \cdot m} b T_0$.
Similarly, a word ending in $(ce)^n d$ can be derived only by
applying the lower part rules $2 \cdot n + 1$ times, leading to a
final derivation part 
$T_0 \stackrel{*}{\ra} (ce)^n T_n \ra (ce)^n d$.
Hence, each derivation of $\omega$ from $\G$ can be decomposed into
an initial and a final part with those properties.

Next, observe that the transitive closure of the
relation $\succ$ on $\NT$, defined by 
\begin{center}
$A \succ B$ 
~ if ~
$A ::= z B$ is an upper part rule for some $z \in \Sigma$,
\end{center}
is asymmetric, i.e.\ an ordering relation.
Therefore, a part of a derivation of $\omega$ from $\G$
that uses only rules from $\G$'s upper part cannot have any
nonterminal repetition.
For a similar reason,
no derivation part using only rules from the lower part
can have any nonterminal repetition.
Hence, the only way a nonterminal repetition can occur in a derivation
of $\omega$ is to repeat a nonterminal from the initial derivation
part in the final part.

There are $2^m$ different initial derivation parts
$S_0 \stackrel{*}{\ra} a^{(n+1) \cdot m} b T_0$.
For each $i=\:,1m\i$, either all of $\:,1n{ \overline{X}_{i\i} }$
but none of $\:,1n{ X_{i\i} }$
occur in an initial derivation part, or vice versa.
Each assignment of the variables $\:,1m{x_\i}$ corresponds
uniquely to an initial derivation part
such that $x_i$ is assigned $1$ iff $\overline{X}_{ij}$ 
occurs in the part but $X_{ij}$ does not,
for $j=\:,1n\i$.

Assume some fixed initial derivation part 
$S_0 \stackrel{*}{\ra} a^{(n+1) \cdot m} b T_0$
has been chosen, corresponding to some fixed truth value assignment to
$\:,1m{x_\i}$.
As 
Fig.~\ref{Satisfied literal vs. repetition-free $ce$ derivation in Lem.}
demonstrates,
a subsequent derivation 
$T_{j-1} \ra c \gamma_{jk} \ra ce T_j$
causes a repetition iff 
the literal $y_{ik}$ in the $j$th conjunct
isn't true in the chosen assignment:
\begin{itemize}
\item Column $y_{jk}$ lists the possible forms that this literal can take,
	where $i$ is choosen such that 
	$y_{jk} \in \set{ x_i, \overline{x_i} }$,
\item column $x_i$ lists the possible truth values assigned to $x_i$,
\item column ``sat'' shows for each possibility whether the literal
	$y_{jk}$ is satisfied (``$+$'') or not (``$-$''),
\item column ``initial'' shows, for each possibility, the nonterminal
	of the initial derivation part
	corresponding to the assignment to $x_i$,
\item column ``final'' shows, for each possibility, the nonterminal
	$\gamma_{jk}$ of the final derivation part 
	$T_{j-1} \ra c \gamma_{jk} \ra ce T_j$,
\item column ``rep'' shows, for each possibility, whether the latter
	nonterminal of the final part is a repetition of that from the
	initial part.
\end{itemize}

Since each possible path $T_{j-1} \ra \ra ce T_j$
involves some $\gamma_{jk}$, each such path 
causes a nonterminal repetition
iff the $j$th conjunct, $y_{j1}+y_{j2}+y_{j3}$, isn't satisfied
by the assignment.

Since the only way to have a repetition is between the initial part
and some $T_{j-1} \ra \ra ce T_j$ part, we have:
Each derivation of $\omega$ starting with
the chosen initial derivation part leads to a repetition iff
the corresponding truth value assignment doesn't satisfy the formula.

Hence, no repetition-free derivation of $\omega$
exists iff the formula is unsatisfiable.
\qed
}

\COROLLARY{Repetition-Free Derivability from a Regular Grammar is NP-Hard}{%
\label{Repetition-Free Derivability from a Regular Grammar is NP-Hard}%
The task to decide whether a given word $\omega$ has a
derivation without nonterminal repetition
from a given regular grammar $\G$
is NP-hard.
}
\PROOF{
Let a conjunctive normal form formula $\kappa$
be given as in Def.~\ref{Conjunctive normal form formula}.
Let $\G$ be the corresponding grammar as in Def.~\ref{Grammar
corresponding to a conjunctive normal form}, let
$\omega = a^{(n+1) \cdot m} b (ce)^n d$.
By Lem.~\ref{Repetition-Free derivability}, the NP-complete
problem to decide whether $\kappa$ is satisfiable
can be reduced to the task to decide whether $\omega$ is derivable
from $\G$ without nonterminal repetition.
\qed
}

\EXAMPLE{Satisfiability and repetion-free derivability}{
\LABEL{Satisfiability and repetion-free derivability}%
Continuing Exm.~\ref{Conjunctive normal form and corresponding grammar},
we consider derivations of the word $\omega = a^{16} b (ce)^3 d$;
this word is derivable in a large number of ways.
Each derivation contains an initial segment like e.g.\
$$\begin{array}{l c rl c rl c rl c rl c rl l}
S_0 
& \ra & a & \0{X_{11}}
	& \ra & aa & \0{X_{12}}
	& \ra & a^3 & \0{X_{13}}
	& \ra & a^4 & S_1	\\
& \ra & a^5 & \3{\overline{X}_{21}}
	& \ra & a^6 & \3{\overline{X}_{22}}
	& \ra & a^7 & \3{\overline{X}_{23}}
	& \ra & a^8 & S_2	\\
& \ra & a^9 & \4{X_{31}}
	& \ra & a^{10} & \4{X_{32}}
	& \ra & a^{11} & \4{X_{33}}
	& \ra & a^{12} & S_3	\\
& \ra & a^{13} & \7{\overline{X}_{41}}
	& \ra & a^{14} & \7{\overline{X}_{42}}
	& \ra & a^{15} & \7{\overline{X}_{43}}
	& \ra & a^{16} & S_4 
	& \ra & a^{16}b & T_0 & ,	\\
\end{array}$$
where for each variable $x_i$ either all nonterminals
$X_{i1}, X_{i2}, X_{i3}$, or all nonterminals
$\overline{X}_{i1}, \overline{X}_{i2}, \overline{X}_{i3}$ occur;
this corresponds to an assignment of $0$ or $1$ to $x_i$.
In our initial segment example, the derivation
corresponds to the assigment $x_1=x_3=0$ and $x_2=x_4=1$.
In a final segment, we have derivations like
$$T_0
	\ra c \0{X_{11}}
	\ra ce T_1
	\ra cec \2{X_{22}}
	\ra (ce)^2 T_2
	\ra (ce)^2c \6{X_{43}}
	\ra (ce)^3 T_3
	\ra (ce)^3 b .
$$
Such a derivation may contain a repetition of
a nonterminal from the initial segment.
In our example,
$T_0 \ra c \0{X_{11}} \ra ce T_1$
contains the repetition of $\0{X_{11}}$,
and correspondingly the propositional variable occurrence
$\0{x_1}$ in the first conjunct is not satisfied by the above assignment.
However,
$T_0 \ra c \6{X_{41}} \ra ce T_1$
does not contain a repetition,
and the first conjunct is satisfied by the
assignment since $\6{x_4}$ is.
\qed
}


\section{Longest repetition-free derivable words}
\LABEL{Longest repetition-free derivable words}

We suspect that the correspondance 
from Def.~\ref{Grammar corresponding to a conjunctive normal form}
between formula $\kappa$ and grammar $\G$,
or a slightly modified version,
can also be used to prove NP-hardness of
the problem of determining the length
of the longest word derivable from a given grammar without repetition.

We already achieved, in 
Lem.~\ref{Upper bound for repetition-free derivable words},
to establish that
no word longer than $\omega$ from 
Lem.~\ref{Repetition-Free derivability},
i.e.\ longer than $(n+1) \cdot (m+2)$ symbols,
can be derived repetition-free from $\G$.

If $\omega$ was the only word of its length that was repetition-free
derivable from $\G$, we had that the longest repetition-free derivable
word has length $(n+1) \cdot (m+2)$ iff $\kappa$ is satisfiable, and a
properly shorter length otherwise.
However, as
Exm.~\ref{Length issues}
shows, there are other words of length $(n+1) \cdot (m+2)$
that are repetition-free derivable from $\G$, but don't correspond to
a truth value assignment in an obvious way.
If we always could construct from such a word a corresponding
satisfying assignment, we had proven the suspected NP-hardness result.

\LEMMA{Upper bound for repetition-free derivable words}{%
\LABEL{Upper bound for repetition-free derivable words}%
No word longer than $(n+1) \cdot (m+2)$ can be derived
repetition-free from the grammar $\G$ from 
Def.~\ref{Grammar corresponding to a conjunctive normal form}.
}
\PROOF{
Let $\psi$ be a word that can be derived repetition-free from $\G$.
First, $\psi$ contains exactly one symbol $d$.
Next, every production of a symbol $b$ or $e$
increases the number of nonterminals from $\set{ \:,0n{T_\i}}$ that
occurred in the derivation,
hence $\psi$ can contain at most $n+1$ such symbols.

We now prove an upper bound on the total number of $a$ and $c$ symbols in
$\psi$.
Assign a pair $\tpl{s^*,j^*}$ to every intermediate word in the
derivation chain of $\psi$, where 
\begin{itemize}
\item $s^*$ 
	is the number of nonterminals from $\set{ \:,0m{S_\i} }$ 
	that already occurred, and 
\item $j^*$ is the current ``conjunction index'',
	i.e.\ 
	\begin{itemize}
	\item $j^*=j$ if the current nonterminal is
		$T_j$ or some $X_{ij}$ or $\overline{X}_{ij}$,
	\item $j^*=0$ if the current nonterminal is some $S_i$, and
	\item $j^*=n$ if the current word doesn't contain a nonterminal.
	\end{itemize}
\end{itemize}

We inspect the grammar rules from Fig.~\ref{Grammar rules in Def.}
to show that the current pair
is properly increased wrt.\ the lexicographical order
whenever a symbol $a$ or $c$ is produced:
\begin{itemize}
\item If $S_{i-1} ::= a X_{i1}$ 
	or $S_{i-1} ::= a \overline{X}_{i1}$ is applied,
	\\
	$s^*$ remains unchanged, 
	while $j^*$ is increased from $0$ to $1$.
\item If or $X_{ij} ::= a X_{i,j+1}$
	or $\overline{X}_{ij} ::= a \overline{X}_{i,j+1}$
	is applied,
	\\
	$s^*$ remains unchanged,
	while $j^*$ is increased from $j$ to $j+1$.
\item If $X_{in} ::= a S_i$ 
	or $\overline{X}_{in} ::= a S_i$ is applied,
	$s^*$ is increased, while $j^*$ is reset to $0$.
\item If $T_{j-1} ::= c \gamma_{jk}$ is applied 
	for some $k \in \set{1,2,3}$,
	\\
	$s^*$ remains unchanged, 
	while $j^*$ is increased from $j-1$ to $j$.
\end{itemize}
The remaining rules don't modify the current pair:
%
\begin{itemize}
\item If $S_m ::= b T_0$ is applied,
	$s^*$ remains unchanged, and $j^*$ remains $0$.
\item If $\gamma_{jk} ::= e T_j$ is applied
	for some $k \in \set{1,2,3}$,
	\\
	$s^*$ remains unchanged, and $j^*$ remains $j$.
\item If $T_n ::= d$ is applied,
	$s^*$ remains unchanged, and $j^*$ remains $n$.
\end{itemize}

Since $S_0$ occurrs in every intermediate word,
we have $1 \leq s^* \leq m+1$ and $0 \leq j^* \leq n$
for every possible pair $\tpl{s^*,j^*}$.
Hence, there are $(m+1) \cdot (n+1)$ possible pairs,
and the current pair can be increased
at most $(m+1) \cdot (n+1) - 1$ times.
Therefore, there are at most that much $a$ and $c$ occurrences in $\psi$.

Summing up, the length of $\psi$ cannot exceed
$1 + n+1 + (m+1) \cdot (n+1) - 1 = (m+2) \cdot (n+1)$ symbols.
\qed
}

\EXAMPLE{Length issues}{
\LABEL{Length issues}%
Continuing Exm.~\ref{Conjunctive normal form and corresponding grammar} 
and~\ref{Satisfiability and repetion-free derivability},
observe that there are repetition-free
derivable words of length $(m+2) \cdot (n+1)$ that are
different from $\omega$ and don't correspond to a truth value assignment.
An examples is
$$\begin{array}{lc rl c rl c rl c rl c rl}
S_0
	&              \ra  & a& \0{X_{11}}
	& \stackrel{*}{\ra} & a^5 & \2{X_{21}}
	& \stackrel{*}{\ra} & a^9 & \4{X_{31}}
	& \stackrel{*}{\ra} & a^{13} & \6{X_{41}}	\\
	& \ra & a^{13} e & T_1
	& \ra & a^{13} ec & \7{\overline{X}_{42}}
	& \ra & a^{13} ece & T_2
	& \ra & a^{13} ecec & \6{X_{43}}	\\
	& \ra & a^{13} ecec a & S_4
	& \ra & a^{13} ecec ab & T_0
	& \ra & a^{13} ecec abc & \3{\overline{X}_{21}}	\\
	& \ra & a^{13} ecec abc a & \3{\overline{X}_{22}}
	& \ra & a^{13} ecec abc aa & \3{\overline{X}_{23}}	\\
	& \ra & a^{13} ecec abc aa e & T_3
	& \ra & \multicolumn{2}{l}{a^{13} ecec abc aa ed}	\\
\end{array}$$
This derivation cannot correspond to a variable
assignment, since it contains e.g.\ both $\2{X_{21}}$ and
$\3{\overline{X}_{21}}$.
By Lem.~\ref{Upper bound for repetition-free derivable words}, no
longer word can be derived from the example grammar.

As a side remark, there are shorter words derivable from $S_0$
without repetition, such as
$$S_0
	\ra a \0{X_{11}}
	\ra ae T_1
	\ra aec \2{X_{22}}
	\ra aece T_2
	\ra aecec \1{\overline{X}_{13}}
	\ra ae(ce)^2 T_3
	\ra ae(ce)^2 d
$$
and
$$S_0
	\ra a \1{\overline{X}_{11}}
	\ra aa \1{\overline{X}_{12}}
	\ra a^3 \1{\overline{X}_{13}}
	\ra a^4e T_3
	\ra a^4ed .
$$
Note that the former derivation also no longer corresponds to a variable
assignment, since it contains both $\0{X_{11}}$ and
$\1{\overline{X}_{13}}$.
When repetitions are allowed,
arbitrarily long words can be derived, e.g.\
$$\begin{array}{l c rl c rl c rl c rl c rl l}
S_0
	& \stackrel{*}{\ra} & a^{16}b & T_0	\\
& \ra & a^{16}bc & \6{X_{41}}
	& \ra & a^{16}bc \; a   & \6{X_{42}}
	& \stackrel{*}{\ra} & a^{16}bc \; a^3 & S_4
	& \stackrel{*}{\ra} & a^{16}bc \; a^3bc & \6{X_{41}}	\\
& \stackrel{*}{\ra} 
	& \multicolumn{5}{l}{a^{16}bc \; (a^3bc)^r \; e(ce)^2 d}\\
\end{array}$$
for any $r \geq 0$.
\qed
}

\begin{figure}
$$\begin{array}{lcl@{\hspace*{0.5cm}}l}
\hline
S_{i-1} & ::= & a \; X_{in}
	& \mbox{for } i=\:,1m\i	\\
S_{i-1} & ::= & a \; \overline{X}_{in}
	& \mbox{for } i=\:,1m\i	\\
X_{ij} & ::= & a \; X_{i,j-1}
	& \mbox{for } i=\:,1m\i \mbox{ and } j=\:,n2\i	\\
\overline{X}_{ij} & ::= & a \; \overline{X}_{i,j-1}
	& \mbox{for } i=\:,1m\i \mbox{ and } j=\:,n2\i	\\
X_{i1} & ::= & a \; S_i & \mbox{for } i=\:,1m\i	\\
\overline{X}_{i1} & ::= & a \; S_i 
	& \mbox{for } i=\:,1m\i	\\
S_m & ::= & b T_0	\\
\hline
T_{j-1} & ::= & c \; \gamma_{jk} 
	& \mbox{for } j=\:,1n\i
	\mbox{ and } k=1,2,3	\\
\gamma_{jk} & ::= & e \; T_j
	& \mbox{for } j=\:,1n\i
	\mbox{ and } k=1,2,3	\\
T_n & ::= & d	\\
\hline
\multicolumn{4}{l}{\mbox{where the mapping } \gamma 
	\mbox{ is defined by}}	\\
\gamma_{jk} & = & X_{ij} & \mbox{for } y_{jk} = x_i	\\
\gamma_{jk} & = & \overline{X}_{ij} 
	& \mbox{for } y_{jk} = \overline{x}_i	\\
\hline
\end{array}$$
\caption{Reversed grammar rules in 
	Def.~\ref{Grammar corresponding to a conjunctive normal form}}
\label{Reversed grammar rules in Def.}
\end{figure}

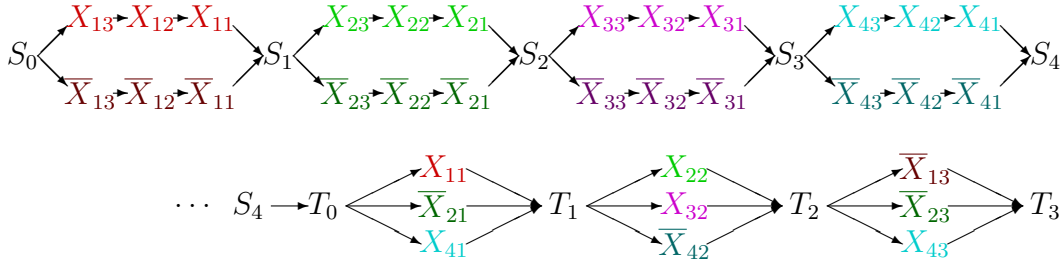
\begin{figure}
\begin{picture}(150,40)
\put(3.000,30.000){\makebox(0.000,0.000){$S_0$}}
\put(5.000,30.000){\vector(1,1){4.000}}
\put(5.000,30.000){\vector(1,-1){4.000}}

\put(12.000,35.000){\makebox(0.000,0.000){$\0{X_{13}}$}}
\put(12.000,25.000){\makebox(0.000,0.000){$\1{\overline{X}_{13}}$}}
\put(15.000,35.000){\vector(1,0){2.000}}
\put(15.000,25.000){\vector(1,0){2.000}}
\put(20.000,35.000){\makebox(0.000,0.000){$\0{X_{12}}$}}
\put(20.000,25.000){\makebox(0.000,0.000){$\1{\overline{X}_{12}}$}}
\put(23.000,35.000){\vector(1,0){2.000}}
\put(23.000,25.000){\vector(1,0){2.000}}
\put(28.000,35.000){\makebox(0.000,0.000){$\0{X_{11}}$}}
\put(28.000,25.000){\makebox(0.000,0.000){$\1{\overline{X}_{11}}$}}
\put(31.000,34.000){\vector(1,-1){4.000}}
\put(31.000,26.000){\vector(1,1){4.000}}

\put(37.000,30.000){\makebox(0.000,0.000){$S_1$}}
\put(39.000,30.000){\vector(1,1){4.000}}
\put(39.000,30.000){\vector(1,-1){4.000}}
\put(46.000,35.000){\makebox(0.000,0.000){$\2{X_{23}}$}}
\put(46.000,25.000){\makebox(0.000,0.000){$\3{\overline{X}_{23}}$}}
\put(49.000,35.000){\vector(1,0){2.000}}
\put(49.000,25.000){\vector(1,0){2.000}}
\put(54.000,35.000){\makebox(0.000,0.000){$\2{X_{22}}$}}
\put(54.000,25.000){\makebox(0.000,0.000){$\3{\overline{X}_{22}}$}}
\put(57.000,35.000){\vector(1,0){2.000}}
\put(57.000,25.000){\vector(1,0){2.000}}
\put(62.000,35.000){\makebox(0.000,0.000){$\2{X_{21}}$}}
\put(62.000,25.000){\makebox(0.000,0.000){$\3{\overline{X}_{21}}$}}
\put(65.000,34.000){\vector(1,-1){4.000}}
\put(65.000,26.000){\vector(1,1){4.000}}

\put(71.000,30.000){\makebox(0.000,0.000){$S_2$}}
\put(73.000,30.000){\vector(1,1){4.000}}
\put(73.000,30.000){\vector(1,-1){4.000}}
\put(80.000,35.000){\makebox(0.000,0.000){$\4{X_{33}}$}}
\put(80.000,25.000){\makebox(0.000,0.000){$\5{\overline{X}_{33}}$}}
\put(83.000,35.000){\vector(1,0){2.000}}
\put(83.000,25.000){\vector(1,0){2.000}}
\put(88.000,35.000){\makebox(0.000,0.000){$\4{X_{32}}$}}
\put(88.000,25.000){\makebox(0.000,0.000){$\5{\overline{X}_{32}}$}}
\put(91.000,35.000){\vector(1,0){2.000}}
\put(91.000,25.000){\vector(1,0){2.000}}
\put(96.000,35.000){\makebox(0.000,0.000){$\4{X_{31}}$}}
\put(96.000,25.000){\makebox(0.000,0.000){$\5{\overline{X}_{31}}$}}
\put(99.000,34.000){\vector(1,-1){4.000}}
\put(99.000,26.000){\vector(1,1){4.000}}

\put(105.000,30.000){\makebox(0.000,0.000){$S_3$}}
\put(107.000,30.000){\vector(1,1){4.000}}
\put(107.000,30.000){\vector(1,-1){4.000}}
\put(114.000,35.000){\makebox(0.000,0.000){$\6{X_{43}}$}}
\put(114.000,25.000){\makebox(0.000,0.000){$\7{\overline{X}_{43}}$}}
\put(117.000,35.000){\vector(1,0){2.000}}
\put(117.000,25.000){\vector(1,0){2.000}}
\put(122.000,35.000){\makebox(0.000,0.000){$\6{X_{42}}$}}
\put(122.000,25.000){\makebox(0.000,0.000){$\7{\overline{X}_{42}}$}}
\put(125.000,35.000){\vector(1,0){2.000}}
\put(125.000,25.000){\vector(1,0){2.000}}
\put(130.000,35.000){\makebox(0.000,0.000){$\6{X_{41}}$}}
\put(130.000,25.000){\makebox(0.000,0.000){$\7{\overline{X}_{41}}$}}
\put(133.000,34.000){\vector(1,-1){4.000}}
\put(133.000,26.000){\vector(1,1){4.000}}

\put(139.000,30.000){\makebox(0.000,0.000){$S_4$}}

\put(28.000,10.000){\makebox(0.000,0.000)[r]{$\ldots$}}
\put(33.000,10.000){\makebox(0.000,0.000){$S_4$}}
\put(36.000,10.000){\vector(1,0){5.000}}
\put(43.000,10.000){\makebox(0.000,0.000){$T_0$}}
\put(46.000,10.000){\vector(2,1){10.000}}
\put(46.000,10.000){\vector(1,0){10.000}}
\put(46.000,10.000){\vector(2,-1){10.000}}
\put(59.000,15.000){\makebox(0.000,0.000){$\0{X_{11}}$}}
\put(59.000,10.000){\makebox(0.000,0.000){$\3{\overline{X}_{21}}$}}
\put(59.000,5.000){\makebox(0.000,0.000){$\6{X_{41}}$}}
\put(62.000,15.000){\vector(2,-1){10.000}}
\put(62.000,10.000){\vector(1,0){10.000}}
\put(62.000,5.000){\vector(2,1){10.000}}
\put(75.000,10.000){\makebox(0.000,0.000){$T_1$}}
\put(78.000,10.000){\vector(2,1){10.000}}
\put(78.000,10.000){\vector(1,0){10.000}}
\put(78.000,10.000){\vector(2,-1){10.000}}
\put(91.000,15.000){\makebox(0.000,0.000){$\2{X_{22}}$}}
\put(91.000,10.000){\makebox(0.000,0.000){$\4{X_{32}}$}}
\put(91.000,5.000){\makebox(0.000,0.000){$\7{\overline{X}_{42}}$}}
\put(94.000,15.000){\vector(2,-1){10.000}}
\put(94.000,10.000){\vector(1,0){10.000}}
\put(94.000,5.000){\vector(2,1){10.000}}
\put(107.000,10.000){\makebox(0.000,0.000){$T_2$}}
\put(110.000,10.000){\vector(2,1){10.000}}
\put(110.000,10.000){\vector(1,0){10.000}}
\put(110.000,10.000){\vector(2,-1){10.000}}
\put(123.000,15.000){\makebox(0.000,0.000){$\1{\overline{X}_{13}}$}}
\put(123.000,10.000){\makebox(0.000,0.000){$\3{\overline{X}_{23}}$}}
\put(123.000,5.000){\makebox(0.000,0.000){$\6{X_{43}}$}}
\put(126.000,15.000){\vector(2,-1){10.000}}
\put(126.000,10.000){\vector(1,0){10.000}}
\put(126.000,5.000){\vector(2,1){10.000}}
\put(139.000,10.000){\makebox(0.000,0.000){$T_3$}}

\end{picture}
\caption{Illustration of reversed example grammar in 
	Exm.~\ref{Conjunctive normal form and corresponding grammar}}
\label{Illustration of reversed example grammar in Exm.}
\end{figure}

In an attempt to remedy the above problems, we modified the grammar
from Def.~\ref{Grammar corresponding to a conjunctive normal form}
as shown in Fig.~\ref{Reversed grammar rules in Def.}.
In the upper part, the $X_{ij}$ are chained in reverse
order, as are the $\overline{X}_{ij}$.
The corresponding example grammar for 
Exm.~\ref{Conjunctive normal form and corresponding grammar}
is illustrated in
Fig.~\ref{Illustration of reversed example grammar in Exm.}.

Almost similar to
Lem.~\ref{Upper bound for repetition-free derivable words},
we established a length upper bound of $(n+1) \cdot (m+2)$ for
repetition-free derivations from the reversed grammar, see
Lem.~\ref{Upper bound for repetition-free derivable words (reversed grammar)}.
The requirement that a word contains a ``b'' symbol could possibly be
overcome if the upper and the lower part were concatenated in reverse
order, i.e.\ by deleting the rules $S_m ::= b T_0$ and $t_n ::= d$,
adding instead the rules $S_m ::= d$ and $T_n ::= b S_0$,
and changing the start symbol to be $T_0$.
However, we didn't elaborate this modification.

\LEMMA{Upper bound for repetition-free derivable words (reversed grammar)}{%
\LABEL{Upper bound for repetition-free derivable words (reversed grammar)}%
For $n \geq 2$,
no word longer than $(n+1) \cdot (m+2)$ and containing a ``b'' symbol
can be derived
repetition-free from the grammar $\G$ from 
Def.~\ref{Grammar corresponding to a conjunctive normal form}.
}
\PROOF{
Let $\psi$ be a word that can be derived repetition-free from $\G$.
Let $a$, $c$, and $e$ denote the number of occurrences of ``a'',
``c'', and ``e'' in $\psi$, respectively.

Assign a ``conjunction index'' to every nonterminal as follows:
\begin{itemize}
\item assign $j$ to each $X_{ij}$, for $i=\:,1m\i$ and $j=\:,1n\i$,
\item assign $n+1$ to each $S_i$, for $i=\:,0m\i$, and
\item assign $j+1$ to each $T_j$, for $j=\:,0n\i$.
\end{itemize}
Observe the following properties:
\begin{itemize}
\item Each increase of the conjunction index in the derivation
	requires some $S_i$ or $T_j$ to occur; 
	neither an occurrence
	of $S_0$ nor one of $T_0$ leads to an increase.
\item More precisely, the conjunction index is increased from $1$ to
	$n+1$ when some $S_i$ occurs,
	and from $j$ to $j+1$ when some $T_j$ occurs.
\item Hence, the conjunction index can experience at most a total
	increase of $mn+n$, if all $m+n$ rules producing a $S_i$ or
	$T_j$ are used.
\item Both the initial and the final conjunction index is $n+1$.
\item Hence the conjunction index'
	total increase must equal the total decrease.
\item 
	If rule $S_m ::= b T_0$ is applied, decreasing the conjunction
	index from $n+1$ to $1$, at most $mn$ ``a''-producing rules
	can be applied, each of them decreasing the conjunction index
	by $1$.
	That is, there are at most $1+mn$ decreasing rule applications.
\item Each grammar rule changes the conjunction index, except where a
	``c'' is produced, by a rule $T_{j-1} ::= c \gamma_{jk}$.
\item Adding up the upper bound for the number of rule applications
	that increase, decrease, and keep the conjunction index,
	and the inevitable final one $T_n ::= d$,
	we get $(m+n) \;+\; (mn+1) \;+\; n \;+\; 1 \;=\; (n+1)(m+2)$.
\qed
\end{itemize}
}

\EXAMPLE{Length issues (reversed grammar)}{
\LABEL{Length issues (reversed grammar)}%
For the reversed grammar scheme,
there are still derivable words of length $(m+2) \cdot (n+1)$ that are
different from $\omega$ and don't correspond to a truth value
assignment.
An example, based on the grammar for 
$(\0{x_1}+\3{\overline{x}_2}+\6{x_4})
\cdot (\0{x_1}+\4{x_3}+\1{\overline{x}_1})
\cdot (\1{\overline{x}_1}+\3{\overline{x}_2}+\2{x_2})$
is the following.
$$\begin{array}{l c rl c rl c rl}
S_0
	&              \ra  & a& \0{X_{13}}
	&              \ra  & aa & \0{X_{12}}
	&              \ra  & aae & T_2	\\
	&              \ra  & aaec & \3{\overline{X}_{23}}
	&              \ra  & aaeca & \3{\overline{X}_{22}}
	&              \ra  & aaecaa & \3{\overline{X}_{21}}	\\
	&              \ra  & aaeca^3 & S_2
	&              \ra  & aaeca^4 & \5{\overline{X}_{33}}
	& \stackrel{*}{\ra} & aaeca^8 & \7{\overline{X}_{43}}	\\
	& \stackrel{*}{\ra} & aaeca^{11} & S_4
	&              \ra  & aaeca^{11}b & T_0
	&              \ra  & aaeca^{11}bc & \0{X_{11}}	\\
	&              \ra  & aaeca^{11}bce & T_1
	&              \ra  & aaeca^{11}bcec & \1{\overline{X}_{12}}
	&              \ra  & aaeca^{11}bceca & \1{\overline{X}_{11}} \\
	&              \ra  & aaeca^{11}bcecaa & S_1
	&              \ra  & aaeca^{11}bceca^3 & \2{X_{23}}
	&              \ra  & aaeca^{11}bceca^3e & T_3	\\
	&              \ra  & aaeca^{11}bceca^3ed	\\
\end{array}$$
Note that the 2nd and 3rd conjunct of the conjunctive normal form
are trivial, as they contain a variable and its negation.
It is not yet clear whether there are similar counter-examples for
non-trivial normal forms.
\qed
}

}


\definecolor{coEv}	{rgb}{0.00,0.99,0.50}
\definecolor{coEe}	{rgb}{0.00,0.50,0.00}
\definecolor{coTm}	{rgb}{0.00,0.00,0.50}

\newcommand{\0}[3]{\textcolor{coEe}{#1#2}\textcolor{coEv}{#3}}
\newcommand{\1}[1]{\textcolor{coTm}{#1}}
\newcommand{\2}[2]{\textcolor{coEe}{#1#2}}

\section{Application to sequence guessing}
\LABEL{Application to sequence guessing}

A modification of
Cor.~\ref{Repetition-Free Derivability from a Regular Grammar is NP-Hard}
can be applied to a problem in artificial intelligence; this was our
original motivation to investigate repetition-free derivations.

One of the typical tasks
in classical intelligence tests is to guess a plausible
construction law for a given sequence of values.
For example, the sequence $0; 2, 4, 6, 8$ has construction laws like
$v_p*2$ and $v_1+2$, where $v_p$ and $v_1$ denotes
the position\footnote{%
	starting with $0$
}
within the sequence and the previous sequence value,\footnote{%
	Since $v_1$ is undefined at position $0$, the first value
	cannot be constructed that way.
	We indicate by a semi-colon the first sequence position where a 
	construction law shall apply.
}
respectively.

Given a sequence $s$ and a set $\Sigma$ of admitted arithmetic operations,
the set of all construction law terms for $s$ that can be
built from $\Sigma$ can be computed as a regular tree
grammar\footnote{%
	an extension of regular word grammars that share
	their closure and decidability properties, while describing
	sets of trees (i.e.\ terms), rather than words; 
	their terminal symbols are function symbols of arbitrary arity;
	see e.g.\
	\cite{Comon.Dauchet.Gilleron.2008}
}
by $E$-generalization\footnote{%
	i.e.\ anti-unification w.r.t.\ an equational background theory
	defining the semantics of operations in $\Sigma$
}
\cite{Heinz.1995},
\cite[Sect.5.2, p.28--29]{Burghardt.2005c}.

As a formalization of Occam's Razor, a law term should be as small as
possible w.r.t.\ some user-definable notion of {\em size\/};
we call such a term {\em guessable\/} from the sequence.
For any reasonable notion of size, a law term\footnote{%
	e.g.\
	$(\mbox{\bf if } v_p<5 \mbox{ \bf then } v_p*2 
	\mbox{ \bf else } 9)$
	for the above example sequence
}
should be discarded if a proper subterm constructs the same sequence, too.
In the grammar setting,
the latter condition amounts to discarding each term whose derivation
uses a nonterminal repeatedly on the same term path.
This is where repetion-free derivations come into play.

Based on our formalization, one may investigate various properties of
a given intelligence test.
Given $\Sigma$, a sequence $s$, and a proper prefix sequence $s'$,
one may e.g.\ ask whether some law term $t$
for $s$ is guessable already from $s'$.\footnote{
	In that case, being asked for a plausible continuation of $s'$, a
	valid answer would be $s$, based on the construction law $t$ as a
	rationale.
	As a counter-example, the term
	$(\mbox{\bf if } v_p<5 \mbox{ \bf then } v_p*2 
	\mbox{ \bf else } 9)$
	is guessable from $0, 2, 4, 6, 8, 9$,
	but from none of its proper prefixes,
	since the subterm $v_p*2$ constructs each of them.
}
Since the law term grammar for $s'$ is a quotient of the grammar
$\G$ for $s$, w.r.t.\ some equivalence relation $\ntEquiv$,
we are searching for a term $t$ whose derivation from $\G$ has no
repetitions w.r.t.\ $\ntEquiv$.

Corollary~\ref{Existence of repetition-free derivations mod. equivalence is NP-hard}
below shows that this search task unfortunately is NP-hard already for
the special case of regular word grammars.\footnote{%
	i.e.\ even when all involved operator symbols are unary or nullary
}
It uses the technical result from 
Lem.~\ref{Repetition-Free derivability}.

Before giving the Corollary, we formalize some of the notions
introduced above.

\DEFINITION{Repetition-free derivation modulo equivalence}{%
\LABEL{Repetition-free derivation modulo equivalence}%
Given a regular grammar $\G'$ and an equivalence relation $\ntEquiv$
on its set $\NT'$ of its nonterminals,
define a derivation from $\G'$ to be repetition-free mod.\
$\ntEquiv$ if it doesn't contain two nonterminals that are equivalent
mod.\ $\ntEquiv$.
\qed
}

\DEFINITION{Quotient grammar}{%
\LABEL{Quotient grammar}
Let $\G' = \tpl{\NT',\Sigma',\R',S'}$ be a regular grammar,
and $\ntEquiv$ be an equivalence relation on $\NT'$.
Similar to the construction of a quotient of a finite
automaton,\footnote{%
	This definition is used in connection with minization of
	deterministic finite automata,
	but often left implicit in textbooks 
	(e.g.\ \cite[Sect.3.4, p.65--71]{Hopcroft.Ullman.1979});
	see e.g.\ \cite[p.5]{Gall.Jeannet.2007} for an explicit
	definition.
}
we can define the quotient grammar $\G = \G'/_\ntEquiv$
of $\G'$ by $\ntEquiv$ to be $\G = \tpl{\NT,\Sigma,\R,S}$, where
\begin{itemize}
\item the nonterminal alphabet $\NT = \NT'/_\ntEquiv$ of $\G$
	is the set of all equivalence classes of nonterminals from $\NT'$,
\item the terminal alphabet $\Sigma = \Sigma'$ of $\G$
	is shared with $\G'$,
\item the rules $\R$ of $\G$ are obtained by replacing all
	nonterminals in all rules in $\R'$ by their equivalence classes,
	and
\item the start symbol $S = S'/_\ntEquiv$ of $\G$
	is the equivalence class of the start symbol of $\G'$.
\end{itemize}

It is obvious that every derivation from $\G'$ can be ``lifted'' to a
derivation from $\G$, by replacing each nonterminal by its equivalence
class.
Hence, $\LL(\G') \subseteq \LL(\G)$, similar to the the well-known
property for quotient automata.
\qed
}

\COROLLARY{Existence of repetition-free derivations mod.\ equivalence
	is NP-hard}{%
\LABEL{Existence of repetition-free derivations mod. equivalence is NP-hard}%
Given a regular grammar $\G'$ and an equivalence relation $\ntEquiv$ on
the set of its nonterminals, the problem to decide whether some
word $\omega \in \LL(\G')$ has a derivation from $\G'$ without repetitions
mod.\ $\ntEquiv$, is NP-hard in general.
}
\PROOF{
Let a conjunctive normal form formula $\kappa$
be given as in Def.~\ref{Conjunctive normal form formula}.

We construct a regular grammar $\G'$ and an equivalence relation
$\ntEquiv$ on its set $\NT'$ of nonterminal symbols such that:
a word $\omega \in \LL(\G')$ exists 
that has a repetition-free derivation mod.\ $\ntEquiv$
iff $\kappa$ has a satisfying variable assignment.

Let
$\NT'
=    \set{ \:,0m{S_\i}, \:,0n{T_\i} }
\cup \set{ X_{ij}, \overline{X}_{ij}, X'_{ij}, \overline{X}'_{ij}
        \mid 1 \leq i \leq m \land 1 \leq j \leq n }$.
Let the rules of $\G'$ be as shown in 
Fig~\ref{Grammar rules in Def.}, except that
the mapping $\gamma$ is now defined as
\begin{itemize}
\item $\gamma_{jk} = X'_{ij}$ for $y_{jk} = x_i$, and 
\item $\gamma_{jk} = \overline{X}'_{ij}$ for $y_{jk} = \overline{x}_i$.
\end{itemize}
Define $(\ntEquiv)$ such that
\begin{itemize}
\item $X_{ij} \ntEquiv X'_{ij}$ for $i=\:,1m\i$ and $j=\:,1n\i$, 
\item $\overline{X}_{ij} \ntEquiv \overline{X}'_{ij}$ 
	for $i=\:,1m\i$ and $j=\:,1n\i$, and
\item no other nontrivial equivalences hold.
\end{itemize}
Observe that the grammar $\G'$ doesn't have any recursion involved, so
its language is finite.
In fact, $\omega = a^{(n+1) \cdot m} b (ce)^n d$ 
from Lem.~\ref{Repetition-Free derivability}
is the only word that can be
derived from $\G'$, but there are lots of different derivations that
accomplish this.
Furthermore, the quotient grammar $\G'/_\ntEquiv$ just
yields the grammar
$\G$ from 
Def.~\ref{Grammar corresponding to a conjunctive normal form}.
Each derivation from $\G'$ corresponds to a derivation from $\G$, but
not vice versa, as observed in Def.~\ref{Quotient grammar}.

A derivation of some word, i.e.\ $\omega$,
from $\G'$ is repetition-free mod.\ $\ntEquiv$
iff
that derivation, taken from $\G$, is repetition-free,
that is,
iff (by Lem.~\ref{Repetition-Free derivability})
$\kappa$ is satisfiable.
\qed
}

Cor.~\ref{Existence of repetition-free derivations mod. equivalence is NP-hard}
subdues our hope to find an efficient algorithm to decide whether a
law term (constructed from a given set of operators)
for a given sequence $s$ is guessable from a given prefix $s'$.

Note, however, that repetition-free derivability mod.\ $\ntEquiv$
is a necessary, but
not sufficient condition for 
$t$ being minimal w.r.t.\ some notion of {\em size\/}.
There are repetition-free (mod.\ $\ntEquiv$)
derivable terms that are nevertheless non-minimal w.r.t.\ every
reasonable notion of {\em size\/}.
For example, $v_p+v_1$ is a construction law term for the sequence
$1; 2, 4, 7$,
none of its subterms is a law for its proper prefix $1; 2, 4$,\footnote{%
	i.e.\ the term $v_p+v_1$ has a repetition-free derivation mod.\
	$\ntEquiv$, where factorizing by the latter turns the grammar
	for $1; 2, 4, 7$ into that for $1; 2, 4$
}
yet every admitted
definition of a {\em size\/} notion will either make
$v_1+v_1$ a smaller or equal term,
or $v_p+v_p$, both are laws for $1; 2, 4$.

As a consequence, the above guessability task could still be
efficiently decidable.

\bibliographystyle{alpha}
\bibliography{lit}

\end{document}